%% file: main.tex
\documentclass[conference]{IEEEtran}
\usepackage{cite}
\usepackage{amsmath,amssymb,amsfonts}
\usepackage{algorithmic}
\usepackage{graphicx}
\usepackage{textcomp}

\usepackage[table, dvipsnames]{xcolor} %for use in color links
\usepackage{soul}
\usepackage{tabularx}
\usepackage{pgfplots, pgf}
\usepackage{pgfplotstable}
\pgfplotsset{compat=1.7}
\usepackage{tikz}
\usepackage{pgf-pie}
\usetikzlibrary{decorations.markings,calc}

\usetikzlibrary{fit}
\usetikzlibrary{backgrounds}
\usetikzlibrary{arrows, petri, topaths}
\pgfdeclarelayer{background}
\pgfdeclarelayer{foreground}
\pgfsetlayers{background,main,foreground}

\usepackage{fancyhdr}
\usepackage[hyphens]{url}

\usepackage[caption=false,font=normalsize,labelfont=sf,textfont=sf]{subfig}

\def\BibTeX{{\rm B\kern-.05em{\sc i\kern-.025em b}\kern-.08em
    T\kern-.1667em\lower.7ex\hbox{E}\kern-.125emX}}

\fancypagestyle{firstpage}{
  \fancyhf{}

  \fancyfoot[C]{\thepage}
}

\title{Exploiting Beam Search Confidence for Energy-Efficient Speech Recognition} 
\author{\IEEEauthorblockN{Dennis Pinto}
\IEEEauthorblockA{\textit{Universitat Politècnica de Catalunya}\\
Barcelona, Spain \\
dpinto@ac.upc.edu}
\and
\IEEEauthorblockN{Jose-María Arnau}
\IEEEauthorblockA{\textit{Universitat Politècnica de Catalunya}\\
Barcelona, Spain \\
jarnau@ac.upc.edu}
\and
\IEEEauthorblockN{Antonio González}
\IEEEauthorblockA{\textit{Universitat Politècnica de Catalunya}\\
Barcelona, Spain \\
antonio@ac.upc.edu}
}
\begin{document}
\maketitle
\thispagestyle{firstpage}
\pagestyle{plain}

%%%%%% -- PAPER CONTENT STARTS-- %%%%%%%%

\input{text/1_abstract.tex}


\input{text/2_introduction.tex}
\input{text/3_background.tex}
\input{text/4_analisis.tex}
\input{text/5_hardware_details}
\input{text/7_methodology}
\input{text/6_experimental_results}
% \fi
%\IEEEpeerreviewmaketitle

\input{text/8_related_work}
\input{text/9_conclusions}

\input{text/_1_acknowledgments}

%%%%%%% -- PAPER CONTENT ENDS -- %%%%%%%%

%%%%%%%%% -- BIB STYLE AND FILE -- %%%%%%%%
\bibliographystyle{IEEEtranS}
\bibliography{bibliography}
%%%%%%%%%%%%%%%%%%%%%%%%%%%%%%%%%%%%

\end{document}

%% file: text/1_abstract.tex
\begin{abstract}
With computers getting more and more powerful and integrated in our daily lives, the focus is increasingly shifting towards more human-friendly interfaces, making Automatic Speech Recognition (ASR) a central player as the ideal means of interaction with machines. Consequently, interest in speech technology has grown in the last few years, with more systems being proposed and higher accuracy levels being achieved, even surpassing \textit{Human Accuracy}. While ASR systems become increasingly powerful, the computational complexity also increases, and the hardware support have to keep pace. In this paper, we propose a technique to improve the energy-efficiency and performance of ASR systems, focusing on low-power hardware for edge devices. We focus on optimizing the DNN-based Acoustic Model evaluation, as we have observed it to be the main bottleneck in state-of-the-art ASR systems, by leveraging run-time information from the Beam Search. By doing so, we reduce energy and execution time of the acoustic model evaluation by $25.6$\% and $25.9$\%, respectively, with negligible accuracy loss.
\end{abstract}

%% file: text/2_introduction.tex
\section{Introduction} \label{sec:introduction}

Nowadays, computers are immensely more powerful than 50 years ago and have become nearly ubiquitous, with more and more people having access to them. However, computer interfaces remain complicated, and most of the time, they require the user to completely focus on the computer in order to use it. Some current trends try to blur the idea of \textit{using a computer}, by proposing systems that interact with humans in a more natural way, with interfaces that are familiar, or directly invisible, to us. In such a world of fluid human-computer interactions, the so-called \textit{Cognitive Computing} is going to be a key component on all kind of systems. 

From all the technologies wrapped under the umbrella term \textit{Cognitive Computing}, \textit{Automatic Speech Recognition} (ASR) will play a key role, becoming the next big leap in human-machine interactions. ASR is a marathon that started around fifty years ago with very high expectations, but soon faced important technology limitations, causing the progress in the field to slow down. However, with more powerful and ubiquitous computers, interest in this technology is again gaining popularity, and the industry and academia are working towards meeting the growing demand. Not long ago an important milestone was reached, when some systems were claimed to perform better than humans in specific conditions for Large Vocabulary Continuous Speech Recognition~\cite{xiong2016achieving}. 

State-of-the-art ASR systems can be classified as either \textit{Hybrid DNN-HMM} or \textit{End-to-End (E2E)} systems. The Hybrid approach consists of a DNN-based \textit{Acoustic Model}, followed by a \textit{Viterbi Beam Search} to generate a word lattice with the most likely transcriptions, that is later re-scored by an RNN-based \textit{Language Model}. To reduce the complexity of the ASR pipeline, the End-to-End solutions aim at generating a transcription from audio features by using a stand-alone DNN. However, E2E systems include a Beam Search with a Language Model (LM) to significantly improve their accuracy, as detailed in section \ref{sec:background}, and therefore, despite being called End-to-End, they usually combine multiple DNNs and a Beam Search to achieve the best recognition accuracy. 

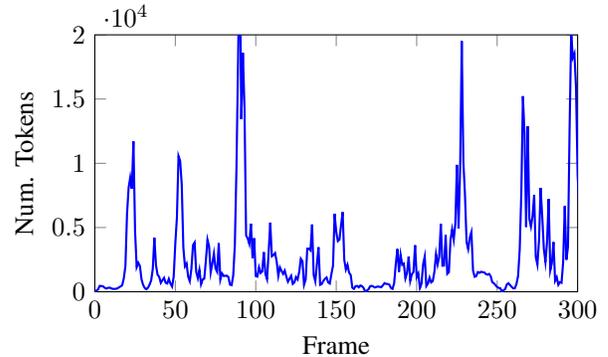
\begin{figure}[!t]
\centering
\input{image/tokensPerFrame.tikz}
\caption{Number of hypotheses, i.e. tokens, expanded by the Beam Search for several frames of speech.}
\label{fig:tokens_per_frame}
\end{figure}

In this work, we aim at improving the energy-efficiency of low-power hardware-accelerated ASR systems. To this end, we first analyze the behavior of a state-of-the-art ASR system when running on an accelerator-rich System-on-Chip (SoC). From the software perspective, the system is representative of recent proposals as it includes a DNN-based Acoustic Scoring stage~\cite{peddinti2015time} and a Beam Search~\cite{rabiner1989tutorial}. From the hardware point of view, our baseline SoC consists of a multicore ARM CPU, a DNN accelerator~\cite{chen2014diannao} and a Beam Search accelerator~\cite{yazdani2016ultra}. Our results show that the DNN evaluation used for acoustic scoring is the main performance and energy bottleneck, consuming $82$\% of the execution time and $68.3$\% of the energy. 

Our key ambition in this work is to improve the energy efficiency of the DNN evaluation, as it is the main performance and energy bottleneck. To this end, we take a novel approach based on the observation that not all the frames require the same level of precision. Figure~\ref{fig:tokens_per_frame} shows the number of hypotheses, a.k.a. tokens, expanded by the Beam Search at each frame of speech. As it can be seen, in some frames the decoder is very confident about the best hypothesis and, hence, the number of tokens expanded is very small, whereas in other parts of the speech the decoder is much less confident, which leads to a large number of hypotheses to be considered. Regions of speech where the Beam Search is not confident require highly accurate DNN scores to avoid discarding the correct hypothesis. However, we show in this work that regions of the speech where the search is highly confident, i.e. the number of tokens is small, do not require a high level of accuracy to make the correct choice and, hence, precision in DNN evaluation can be relaxed with negligible impact in WER.

In this paper, we present a novel technique to reduce the time and energy consumed during ASR, focusing on local real-time evaluation in low-power devices. In this technique, we leverage feedback information from the Beam Search accelerator to select the appropriate accuracy in the DNN accelerator. More specifically, for regions of the speech where the search is highly confident, the DNN is evaluated at low precision (e.g. 4 bits), whereas regions of the speech with low confidence are evaluated at high precision (e.g. 8 bits). By counting the number of tokens at each frame in the Beam Search accelerator, and comparing it with a threshold, we can decide if the level of confidence is high enough to evaluate the Acoustic Model in low precision. The threshold is computed at run-time so it automatically adapts to the particular situation. By using the number of tokens as metric of confidence and the run-time threshold, we can save $16.9$ \% of the energy and reduce the overall execution time by $19.6$\% of the complete system.

This paper focuses on energy-efficient hardware-accelerated ASR. We claim the following contributions:
\begin{itemize}
    \item We characterize the performance and energy consumption of state-of-the-art ASR systems on an accelerator-rich SoC. We identify the evaluation of the DNN-based Acoustic Model as the main bottleneck, as it requires $82$\% and $68.3$\% of the total execution time and energy, respectively.
    \item We analyze the impact of reducing the precision of the DNN-based Acoustic Model in different regions of the speech. We conclude that frames with a high confidence in the search, i.e. with a small number of tokens expanded, can be evaluated at
    low precision with negligible impact in WER.
    \item We present a novel technique that dynamically selects the precision of the DNN accelerator based on the feedback information provided by the Beam Search accelerator. Our system evaluates the DNN at low precision for 50\% of the frames,
    resulting in $19.6$\% reduction in execution time and $16.9$\% energy savings for the entire Librispeech test set.
\end{itemize}

The rest of the paper is organized as follows. Section~\ref{sec:background} contains some background about the software and hardware solutions employed on ASR. In Section~\ref{sec:analysis}, we present our analysis of the ASR system with a discussion of its bottlenecks and a description of the proposed solutions to alleviate them. Section~\ref{sec:platform} is a detailed description of the changes required in the hardware platform to support our technique. In Section~\ref{sec:methodology} we describe our experimental methodology, providing some details regarding the configuration of the system and the models employed for its evaluation and Section~\ref{sec:experimental_results} contains our experimental results. Finally, Sections~\ref{sec:related_work}~and~\ref{sec:conclusions} contain a brief review of the related work and the main conclusions of this work, respectively.

%% file: image/tokensPerFrame.tikz
\begin{tikzpicture}
\begin{axis} [
    height = 5cm,
    width = 8cm,
    xmin = 0,
    xmax = 300,
    ymin = 0,
    ymax = 20000,
    xlabel = Frame,
    ylabel = Num. Tokens,
    %width=\textwidth, 
    %height=0.6\textheight,
    legend style={at={(0.99,0.8)},anchor=east, cells={anchor=west}}
]

\addplot [mark=none, blue, thick] table [x=frame, y=numTokens, col sep=comma] {data/num_tokens_trace.txt};

\end{axis}
\end{tikzpicture}

%% file: text/3_background.tex
\section{State-of-the-art ASR} \label{sec:background}

This section summarizes the state-of-the-art software and hardware solutions for ASR. First, we review the software pipeline of a modern ASR system, that is largely influenced by recent advances in deep learning. Modern ASR systems combine several DNNs used for different sub-tasks, such as Acoustic Scoring or Language Model re-scoring, with other algorithms for speaker adaptation, \textit{Beam Search}, etc. Second, we describe a mobile hardware platform for real-time large vocabulary ASR, including a quad-core ARM CPU integrated with recently proposed accelerators for the most computationally demanding tasks in ASR systems.

\subsection{ASR Pipeline}\label{sec:asr_pipeline}

\begin{figure}[t!]
    \centering
    \includegraphics[scale=0.42]{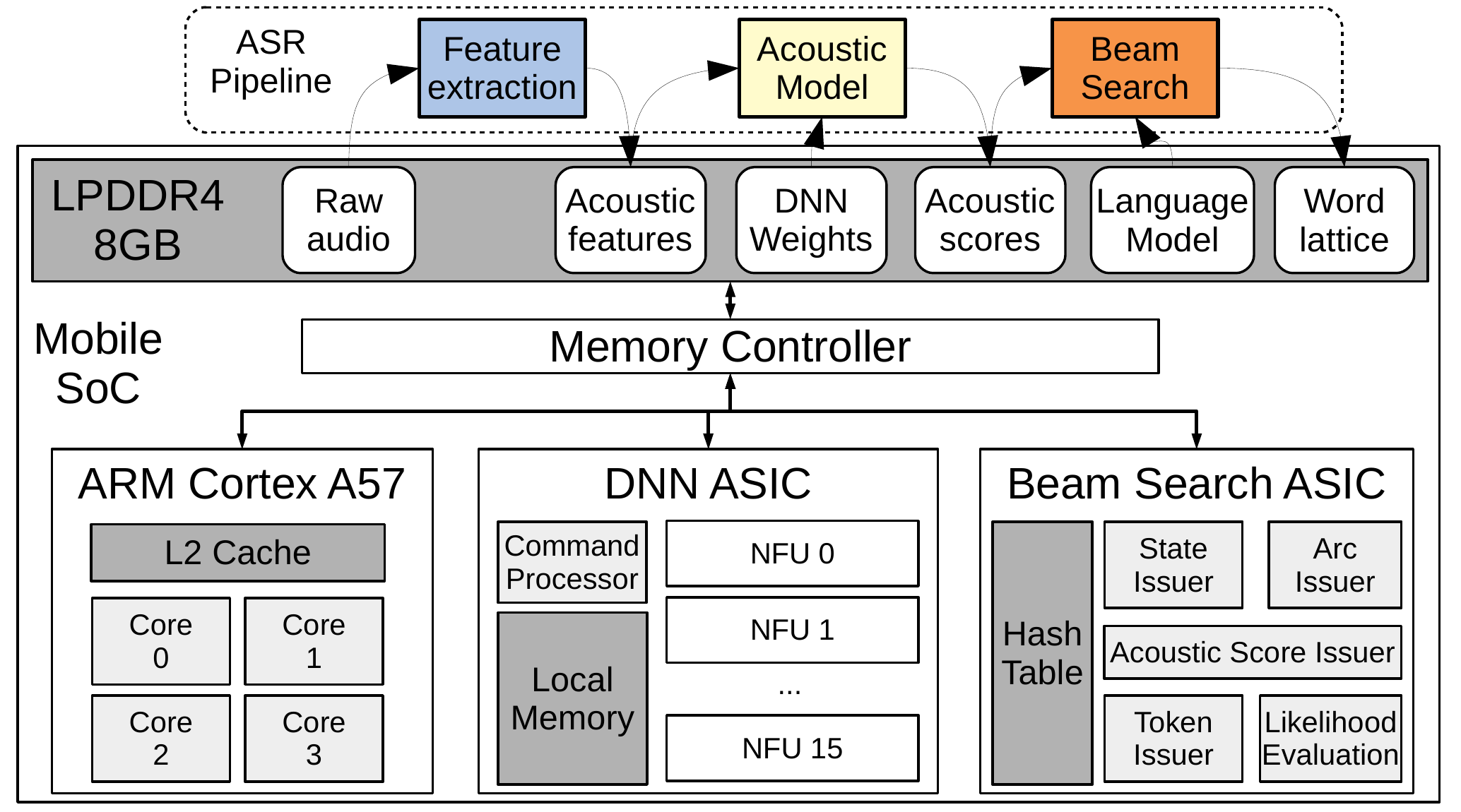}
    \caption{ASR software pipeline running on a mobile SoC for real-time large vocabulary ASR.}
    \label{fig:asr_pipeline}
\end{figure}

The top part of Figure~\ref{fig:asr_pipeline} shows the software pipeline of a modern ASR system. The first stage is the \textit{Feature Extraction}. This component first splits the raw audio signal in overlapping frames of 25 ms of speech. Next, it computes a vector of features to encode each frame, typically \textit{Mel Frequency Cepstral Coefficients}~\cite{logan2000mel} (MFCC). Furthermore, it also computes an \textit{iVector}~\cite{karafiat2011ivector} for speaker adaptation. Therefore, each frame is encoded as a vector that is the concatenation of the audio features and the iVector. The objective of this representation is to expose the information that is relevant for the system in a compact manner.

The next pipeline stage is the evaluation of the Acoustic Model. This model is executed for every frame of speech. It takes as input the acoustic features, i.e. MFCC+iVector, of one frame of speech and its neighbor frames and it computes the probability of that frame representing each of the possible sound units (referred to as sub-phonemes, or senones\footnote{These sound units, sub-phoneme units, or senones, are normally defined as context dependant phonemes, being the most popular the tri-phone. A tri-phone is composed of a central phoneme $p$ plus a left and a right phonemes $p_{l}-p-p_{r}$.}) included in the model.

The state-of-the-art solution to implement the Acoustic Model consists of a DNN trained to compute senones' probabilities, a.k.a. acoustic scores, for each input frame. Although different deep learning solutions, such as Multi-Layer Perceptrons (MLP) or Recurrent Neural Networks (RNN), have been successfully applied for acoustic scoring, a more effective approach is to use a \textit{Time Delay Neural Network} (TDNN)~\cite{peddinti2015time}. Although the use of TDNNs for speech recognition was proposed long ago~\cite{waibel1995phoneme}, they have been proven recently to improve the accuracy of state-of-the-art systems more efficiently than other approaches, such as using RNNs or introducing long contexts to an MLP.

In an E2E system, a greedy algorithm could be used to obtain a transcription from the acoustic scores computed by the Acoustic Model: it just selects the sub-phoneme with maximum score for each frame of speech and performs a simple post-processing to provide a transcription. However, this greedy decoding leads to sub-optimal WER, and accuracy is substantially improved by incorporating a lexicon and a language model by means of a beam search.
For example, Google's E2E system LAS (Listen Attend and Spell) reduces its Word Error Rate (WER) from 6.8\% to 5.8\% in the Librispeech dataset~\cite{panayotov2015librispeech} when using a Beam Search with an LM~\cite{park2019specaugment}. NVIDIA's Jasper improves its WER from 11.9\% to 8.7\% for the same dataset when using a Beam Search and a TransformerXL LM to re-score the likelihoods of the E2E DNN~\cite{li2019jasper}. Baidu's DeepSpeech obtains a relative improvement of 41.6\% in WER when using the output likelihoods of the E2E RNN to drive a Beam Search with a 3-gram LM~\cite{amodei2016deep}. Finally, recent work from Facebook in E2E systems~\cite{synnaeve2019end} also reports large improvements in WER when using a Beam Search and LM re-scoring. Therefore, the state-of-the-art solution is to use the DNN-computed acoustic scores to perform a \textit{Beam Search}. 

The objective of the Beam Search is to find the best path in a graph that contains all the possible sequences for the transcription, weighted by probabilities. This graph is known as the \textit{Decoding Graph} and is usually generated by combining a language model with a Hidden Markov Model and a lexicon. All these models are efficiently represented through a \textit{Weighted Finite State Transducer} (WFST)~\cite{mohri2002weighted}, a mathematical framework to build graphs specifically for sequence-to-sequence translation. The WFST is a type of weighted graph over which there are defined operations that allow to efficiently combine information from several models into the same graph. The use of a WFST makes it possible to merge together a lexicon and a language model into the same graph. 

In order to efficiently traverse the decoding graph, an algorithm known as \textit{Viterbi Beam Search}~\cite{rabiner1989tutorial} is used. This algorithm obtains the less costly path, i.e. the most likely transcription, given the acoustic scores and the WFST-based decoding graph. This Beam Search algorithm works by expanding, for each frame, all the active states in the graph, named \textit{tokens}. At start time, the only token in the \textit{Active Set} represents a special state from the graph, denoted as \textit{Start State}. For each frame of speech, each token in the \textit{Active Set} is replaced by a set of new tokens corresponding to all reachable states from it, each with a weight equal to the addition of the source token weight, the traversed arc's weight and the score obtained by the Acoustic Model for the destination state.

In order to keep the search space manageable, a \textit{Beam Width} is used to discard very unlikely tokens. Hence, after expanding the tokens for a given frame, a pruning step is performed: for each token, the distance between its weight and the best token's weight is computed and if it is larger than the \textit{Beam Width} the token is discarded. Note that due to this pruning the number of alternative tokens considered for each frame may largely vary as shown in Figure~\ref{fig:tokens_per_frame}. More specifically, for regions of speech where the decoder is highly confident only a few tokens are expanded, whereas the Beam Search expands a large number of tokens when it is less confident about the correct transcription. In this work, we exploit the degree of confidence in the Beam Search to perform a more efficient Acoustic Model evaluation: we argue that precision of acoustic scores is critical when the Beam Search shows low confidence, but it can be relaxed when the decoder is highly confident.

The output of the Beam Search is a word lattice containing the most likely transcriptions for the input audio signal. At this point, the best path in the lattice could be recovered by following back-pointers with an inexpensive backtracking step. However, to further improve the WER, the word lattice is normally re-scored by using a more sophisticated language model encoded with an RNN or a Transformer~Network~\cite{synnaeve2019end}.

Note that although our baseline system is a TDNN-based hybrid scheme, the ideas presented in this paper can be applied to any ASR system as long as it contains a DNN-based acoustic model capable of streaming evaluation and a beam search, which includes the vast majority of systems present in the literature.

\subsection{Low-Power Hardware for ASR}\label{sec:asr_hardware}

Figure~\ref{fig:asr_pipeline} illustrates the mobile SoC assumed in this work. It consists of a 
quad-core ARM CPU, a DNN accelerator and a \textit{Beam Search} accelerator, all of them sharing the same system
memory (8GB of LPDDR4). The CPU, an ARM Cortex A57, orchestrates the execution of the ASR pipeline:
it prepares the datasets in main memory and issues commands to the accelerators to offload the most
computationally intensive parts. Furthermore, it also runs part of the \textit{Feature Extraction}.

The DNN accelerator is inspired in DianNao~\cite{chen2014diannao}. It consists of a command processor, local
scratchpad memories and 16 Neural Function Units (NFU). The NFU contains all the units required to perform the 
DNN computations, including an array of adders and multipliers, in addition to specialized units for the 
activation functions. The NFU is pipelined in three stages: NFU-1, to multiply the inputs by the weights;
NFU-2, to add-reduce the results from NFU-1; and NFU-3, to perform the activation function. The internal 
memory is composed of three SRAM buffers to store weights (16KB), inputs (1KB) and outputs (1KB).

We employ the Beam Search accelerator presented in~\cite{yazdani2016ultra}. It contains specialized units to fetch from memory the required data to traverse the decoding graph: state issuer, arc issuer and acoustic likelihood
issuer. The likelihood evaluation unit computes the weights of the new tokens by combining source token weight, the arc's weight and the DNN-computed acoustic score. Finally, the token issuer maintains the list of tokens and generates
the word lattice. 

Regarding the on-chip memories, it includes several caches to speedup the accesses to different data
(state cache of 128KB, arc cache of 256KB and token cache of 128KB) and two hash tables to track the tokens for the current frame and next frame of speech (768KB).

\begin{figure}[!t]
\centering
\includegraphics{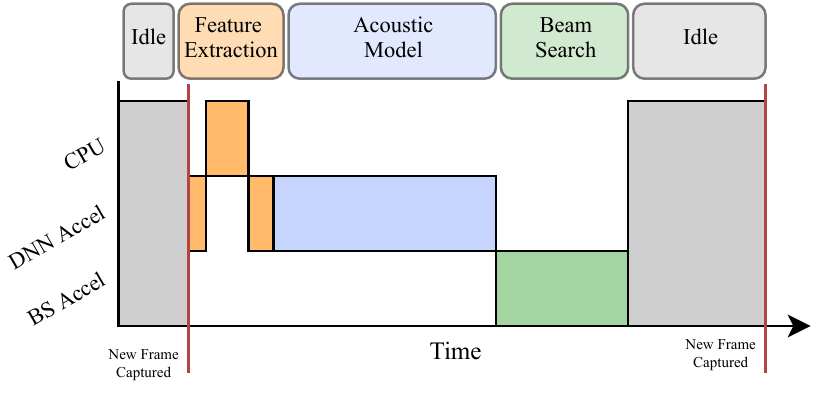}
\caption{Execution of the ASR pipeline on the available hardware. The horizontal axis shows time, whereas the y axis contains a row for each hardware component. When a frame is captured, the features are extracted using the CPU and the DNN accelerator. After that, the \textit{Acoustic Scores} are obtained by evaluating the \textit{Acoustic Model} in the \textit{DNN Accelerator}. These \textit{Acoustic Scores} are then merged with the \textit{Language Model Scores} during \textit{Beam Search}, executed on the \textit{Beam Search Accelerator}. The system is in idle state until a new frame arrives.}
\label{fig:hardware_time_occupation}
\end{figure}

Each ASR component is executed on the best suited hardware. Figure \ref{fig:hardware_time_occupation} shows a diagram of the execution of the different components of the ASR pipeline. The extraction of the features from the audio frames is essentially composed of matrix-vector and matrix-matrix operations, so it can be executed almost exclusively in the DNN accelerator. The extraction of the iVector contains matrix-matrix operations, which can be efficiently handled by the DNN accelerator, and other operations not well-suited for DNN hardware that are executed in the CPU. The Acoustic Model DNN inference and Beam Search are computed in the DNN and Beam Search accelerators, respectively.

Although some steps in the ASR pipeline run on different hardware, they cannot be executed in parallel because each step depends on the previous one. Apart from that, we are assuming a \textit{Stream Evaluation} of the input utterance, in which the frames are evaluated one by one, while the utterance is being captured by the ASR front-end; this is commonly referred to as online real-time speech recognition. Because of using an online ASR system on a hardware that runs faster than real-time, we cannot parallelize among different frames, either. 

Another limitation for parallel execution of the Beam Search and the DNN accelerators is that the DNN evaluation requires a large amount of data from Main Memory. In fact, performance is mainly limited by the amount of bandwidth that it requires.

%% file: text/4_analisis.tex
\section{Analysis of Bottlenecks} \label{sec:analysis}

In this section we analyze the main performance and energy bottlenecks of a state-of-the-art hardware-accelerated ASR system for mobile devices. We implement the ASR pipeline described in Section~\ref{sec:asr_pipeline} using Kaldi~\cite{povey2011kaldi}, a widely used framework for building speech recognition systems. Table~\ref{tbl:table_asr}
shows the relevant parameters of the system. 

The Acoustic Model is a TDNN network, trained to receive as input a 40-dimension MFCC array concatenated with a 100-dimension iVector, complemented with a context of 21 past frames and 21 future frames. The network weights and inputs are linearly quantized to $8$ bits. The output of the Acoustic Model represents the probabilities for the 6056 sub-phoneme elements or \textit{senones} present in the model. The Decoding Graph is a WFST that combines a lexicon of 200k words and a 4-gram language model.

On the other hand, we execute the ASR system in the hardware platform described in Section~\ref{sec:asr_hardware}. Table~\ref{tbl:table_soc} shows the parameters for the mobile SoC. We chose a low-power ARM CPU with 8GB of LPDDR4 DRAM, complemented with two accelerators: one to perform the Beam Search~\cite{yazdani2016ultra} and the other for the DNN inference~\cite{chen2014diannao}. All these subsystems share the same address space, are connected to the main system bus and are served by the same memory controller. The methodology to obtain execution time and energy consumption of this platform is described in Section~\ref{sec:methodology}.

\begin{scriptsize}
\begin{table}[t!]
    \renewcommand{\arraystretch}{1.3}
    \centering
    \rowcolors{1}{lightgray}{white}
    \begin{tabular}{r p{5cm}}
     Acoustic Features & {MFCC (40 elements) + iVector (100 elements)} \\
     TDNN & 26 fully connected layers (16MB)\\
     WFST & 200k words lexicon + 4-gram LM (180MB)\\
    \end{tabular}
    \caption{ASR system parameters}
    \label{tbl:table_asr}
\end{table}
\end{scriptsize}

\begin{table}[!t]
    \renewcommand{\arraystretch}{1.3}
    \centering
    \rowcolors{1}{lightgray}{white}
    \begin{tabular}{r p{5cm}}
    \multicolumn{2}{c}{\textbf{ARM Cortex A57}}\\
     Frequency & 1.7 GHz\\
     Number of cores & 4\\
     Caches & 32KB (L1I), 48KB (L1D), 2MB (L2)\\
     \multicolumn{2}{c}{\textbf{DNN Accelerator}}\\
     Frequency & 55 MHz\\
     SRAM Buffers & 1KB (Input), 1KB (Output), 16KB (Weights)\\
     NFUs & 16 (16 MULs and 16 ADDs per NFU)\\
     \multicolumn{2}{c}{\textbf{Beam Search Accelerator}}\\
     Frequency & 600 MHz\\
     Caches & 128KB (State), 256KB (Arc), 128KB (Token)\\
     Hash Tables & 768KB, 32K entries\\
     Likelihood Evaluation & 4 FP adders and 2 FP comparators\\
     \multicolumn{2}{c}{\textbf{LPDDR4 Main Memory}}\\
     Capacity & 8GB\\
     Bandwidth & 16 GB/s\\
    \end{tabular}
    \caption{Mobile SoC parameters. Technology node is 28nm.}
    \label{tbl:table_soc}
\end{table}

\newpage
\subsection{Bottlenecks}

\begin{figure}[!t]
    \centering
    \centering
    \subfloat[][]{\input{image/energyBreakdown_ASR.tikz}\label{fig:energy_breakdown_ASR}}
    \subfloat[][]{\input{image/energyBreakdown_DNN.tikz}\label{fig:energy_breakdown_DNN}}
    \caption{Breakdown for energy consumption during ASR evaluation on the mobile SoC presented in Section~\ref{sec:asr_hardware}. Chart \textit{(a)} shows the energy breakdown by ASR component, where the clear bottleneck is the Acoustic Model TDNN evaluation, whereas chart \textit{(b)} shows the energy breakdown among hardware components during the AM evaluation. Here, it can be seen how reads and writes from the DRAM are responsible for most of the consumed energy.}%
    \label{fig:energy_breakdown}%
\end{figure}
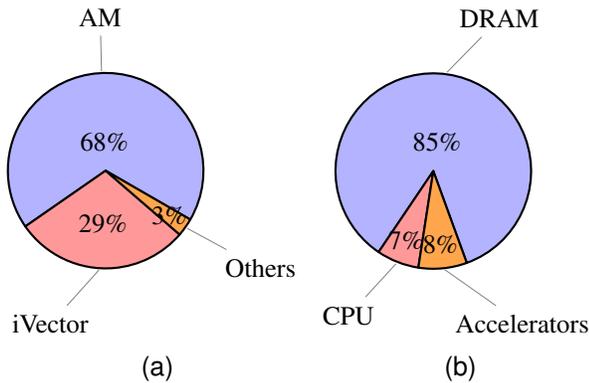

According to our experiments, the distribution of energy consumption among the ASR system components is as shown on Figure~\ref{fig:energy_breakdown}, where we can see how most of it is consumed by the DRAM memory during the DNN inference. Chart~\ref{fig:energy_breakdown_ASR} shows the breakdown of the energy consumed during the evaluation of the \textit{Librispeech} test set, that consists of more than five hours of speech of a large number of speakers, split by the components of the ASR pipeline. The most expensive parts are the iVector and the Acoustic Model evaluation, consuming $28.9$\% and $68.3$\%, respectively. The iVector computation is expensive in terms of energy consumption because it has to be partially executed on the CPU, running at an average power of $2.7 W$. Since the Acoustic Model evaluation is the main bottleneck, our target is to identify how much of that energy is consumed by each hardware component. Chart~\ref{fig:energy_breakdown_DNN} shows the breakdown of the energy consumed during the Acoustic Model evaluation by the different hardware components. The DRAM memory is the clear bottleneck, consuming $85$\% of the energy. The rest of the energy is consumed by the CPU, which is idle during the TDNN evaluation, and the accelerators, of which the Beam Search accelerator is also idle. It is not shut off because it is beneficial to reuse its internal caches and scratchpads across consecutive frames.

Regarding the execution time, TDNN evaluation for acoustic scoring is also the main bottleneck as it takes $82$\% of the execution time. The \textit{Feature Extraction} and the \textit{Beam Search}
require $14.8$\% and $3.2$\% of the execution time respectively. The mobile SoC platform meets the real-time constraints for
all the utterances in Librispeech test set (more than 2k utterances and more than 5 hours of speech), achieving
real-time factors of 0.05xRT on average and 0.09xRT in the worst case.

In conclusion, it is clear from these results that the TDNN is the main performance and energy bottleneck. Furthemore, we have identified that most of the energy ($58.1$\%) is consumed because of main memory accesses during TDNN evaluation. We have measured that 99\% of the memory accesses during TDNN evaluation are accesses for fetching weights. The network is too large to be kept in on-chip memory, and since we are evaluating the input frames one-by-one, we can not take advantage of temporal locality of weights for different input frames.

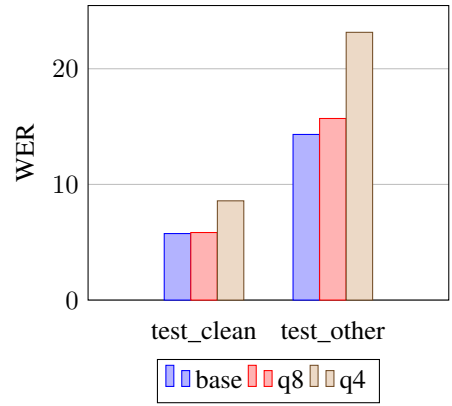
\begin{figure}[!t]
\centering
\input{image/wer-quant_tradeoff.tikz}
\caption{Comparison of the WER loss respect to the non-quantized model for various levels of quantization. While quantizing to $8$ bits has a small impact on WER, more aggressive quantization sensibly degrades accuracy}
\label{fig:wer-quant_tradeoff}
\end{figure}

A well-known optimization to alleviate this bottleneck consist on pruning the network by an iterative process of removing some weights and retraining. This approach results in a sparse network, which can be significantly smaller than the dense network. However, the pruning algorithm is expensive, and the inference with a sparse network requires important changes in the DNN accelerator.

Another well-known technique to alleviate this bottleneck is aggressive DNN quantization. However, quantizing to less than $8$ bits results in an important degradation in recognition accuracy, making it a bad solution. Figure~\ref{fig:wer-quant_tradeoff} shows the WER for \textit{test\_clean} and \textit{test\_other} evaluated with the TDNN acoustic model at different levels of quantization. While $8$ bits results in minor accuracy loss compared to full precision, going to $4$ bits increases the WER by $49$\% in test\_clean and $61$\% in test\_other, and if the weights are quantized to $2$ bits, the system does not work, generating invalid transcriptions. 

In the following sections, we show how Beam Search confidence can be leveraged to reduce energy consumption and increase performance by dynamically selecting the appropriate precision for the Acoustic Model inference.

%% file: image/energyBreakdown_ASR.tikz
\begin{tikzpicture}
 
    \pie[radius=1.3, text=pin, rotate=330, color={blue!30, red!40, orange!70}]  
        {68/AM, 29/iVector, 3/Others}
        
 \end{tikzpicture}
       

%% file: image/energyBreakdown_DNN.tikz
\begin{tikzpicture}

    \pie[radius=1.3, text=pin, rotate=290, color={blue!30, red!40, orange!70}]  
        {85/DRAM, 7/CPU, 8/Accelerators}
        
 \end{tikzpicture}
       

%% file: image/wer-quant_tradeoff.tikz
% Preamble: \pgfplotsset{width=7cm,compat=1.5.1}
\begin{tikzpicture}
    \begin{axis}[
    height = 5.5cm,
    major x tick style = transparent,
    ybar=0.05pt,
    enlarge x limits=0.9,
    bar width=10pt,
    ymajorgrids = true,
    ylabel={WER},
    %symbolic x coords={base,q8,q4},
    symbolic x coords={test\_clean, test\_other},
    xtick = data,
    scaled y ticks = false,
    ymin=0,
    ytick style={draw=none},
    legend cell align=left,
    legend style={at={(0.5,-0.2)},
            anchor=north,
            legend columns=-1
            },
    %nodes near coords,
    %nodes near coords align={vertical},
    ]
        
        \addplot coordinates {(test\_clean,5.75) (test\_other,14.32)};
        \addplot coordinates {(test\_clean,5.84) (test\_other,15.7)};
        \addplot coordinates {(test\_clean,8.58) (test\_other,23.15)};
        
        \legend{base, q8, q4}
        
    \end{axis}
\end{tikzpicture}

%% file: text/5_hardware_details.tex
\section{Dynamic DNN Precision} \label{sec:platform}

This section describes our technique to dynamically set the precision of the DNN used for acoustic scoring based on the confidence of the Beam Search. In Section~\ref{sec:analysis} we showed that fetching the DNN weights from main memory takes a large percentage of the total energy consumption. Furthermore, we showed that reducing DNN precision for all the frames results in a significant accuracy loss. Therefore, we take a different approach and propose to reduce the precision for parts of the speech where the confidence of the decoder is high and, hence, it does not require highly accurate acoustic scores to avoid discarding the correct hypothesis.

\begin{figure}[!t]
    \centering
    \centering
    \subfloat[][]{\input{image/comparison_heuristics_test_clean.tikz}\label{fig:comparison_heuristics_clean}}
    %\qquad
    \subfloat[][]{\input{image/comparison_heuristics_test_other.tikz}\label{fig:comparison_heuristics_other}}
    \caption{WER for a varying percentage of frames evaluated at low precision for a) test\_clean and b) test\_other. The curves represent the cases when the frames for low-precision evaluation are those with less number of tokens (high confidence), and when they are chosen randomly.}
\label{fig:comparison_heuristics}%
\end{figure}
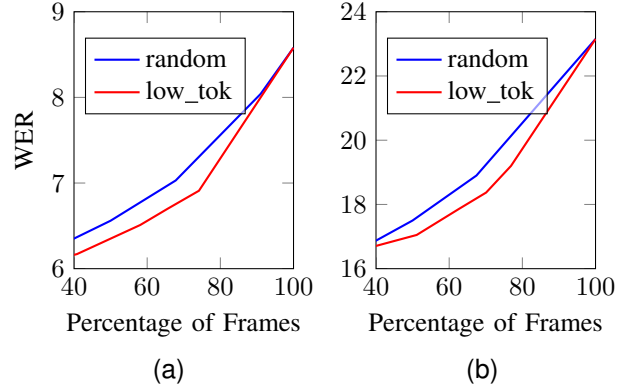

Determining in advance which frames require more or less precision is a challenging problem. To solve that, we propose to look at the number of tokens expanded during the Beam Search, based on the observation that a high number of tokens means that the Beam Search is not very confident about the correct transcription, whereas a low number of tokens is related with a high confidence. Figure~\ref{fig:comparison_heuristics} compares the WER obtained for test\_clear and test\_other with different percentages of frames computed at \textit{4 bits}. When the precision is reduced for frames with a low number of tokens, we can evaluate in low precision many more frames while maintaining a low WER loss, than if the frames are chosen randomly. Therefore, frames with high confidence, i.e. low number of expanded tokens, are good candidates to be evaluated in low precision in the DNN-based acoustic model. Note that when the DNN has to be evaluated, the number of tokens is already known by the Beam Search accelerator, and thus, this information can be exposed to the DNN accelerator to set the precision at run-time.

\subsection{Threshold Computation}
Our technique decides which frames of speech show high/low confidence in the decoding, based on the number of tokens expanded during the previous Beam Search step. If the evaluation of one frame results on few current hypotheses, the DNN inference for the next frame is performed in low precision. The threshold for the number of tokens is set dynamically with the goal of trying to achieve a given target percentage of frames evaluated at low precision. There are two options to set this ratio, a per-utterance ratio, meaning that each utterance will be forced to keep it; or a global ratio, meaning that the ratio will not be forced for each utterance but it will be achieved after computing a number of utterances.
We have observed that a global ratio gives better results because different utterances have different precision requirements. If the threshold is set individually for each utterance, we are undermining that, and so the global WER increases.

The easiest solution for a global threshold would be to measure the number of tokens for each frame in the test set, and fix the threshold according to the desired ratio. However, we have observed that the threshold obtained with this method does not match the desired ratio of low precision frames in the test sets (Figure~\ref{fig:discrepance_thres_simple}). In this figure, the train set was evaluated with high precision, then we sorted the frames according to the number of tokens expanded by Beam Search, and selected the frames at percentiles $30$, $50$ and $70$. The number of tokens at those frames was used as threshold to evaluate test\_clear and test\_other and measure the ratio of frames classified for low precision. This approach to set the threshold results in a percentage of low precision frames significantly lower than in the train set. The reason is that when low precision is used, the number of tokens expanded during the Beam Search changes with respect to the case when high precision is always used, modifying the token distribution and moving the percentiles. 

Figure~\ref{fig:num_tokens_cum_freq} shows the cumulative frequency of frames from test\_clear and test\_other according to the number of tokens expanded, when the DNN is evaluated completely at $8$ bits and $4$ bits. In this case, the percentile $50$ for the test set evaluated at $8$ bits is at $883$ tokens, whereas when the test set is evaluated at $4$~bits, it is at $1961$ tokens. It is clear from the figure that when the acoustic model is evaluated at lower precision, the number of tokens per frame increases. In other words, the overall confidence decreases when low precision is used.

We propose instead an alternative heuristic that tries to compute $50$\% of the frames at low precision by doing run-time adjustments to the threshold. For that, we keep a variable, $h$, that contains the difference between the number of frames evaluated in low precision and those computed at high precision, and try to keep it at $0$, by increasing or decreasing the threshold for the number of tokens.

To avoid the threshold from oscillating wildly, we define an additional variable, $h_l$, which also contains the difference between high precision and low precision frames, but constrained to a window of latest frames. This way, we know if we need more frames at low or high precision, and also the current tendency. If we have more low-precision than high precision frames in the local window ($h_l > 0$), we assume that the number of frames at low precision is increasing. If the opposite is true, i.e. $h_l < 0$, we assume that the number of frames evaluated at low precision is decreasing. With these values, we update the threshold ($Th$) using the following formula:

\begin{equation}
    Th = 
        \begin{cases}
            Th - \Delta & h > 0 \And h_l > 0 \\
            Th + \Delta &  h < 0 \And h_l < 0 \\
        \end{cases}
\end{equation}

When the system has evaluated more low-precision than high-precision frames and the tendency goes towards increasing low-precision frames, the heuristic decreases the threshold by $\Delta$, so it is more difficult for frames to be classified for low precision. On the other hand, if it has evaluated less frames in low precision, both globally and in the local window, the threshold is increased, so more frames are classified as low-precision. Both $\Delta$ and the starting value of $Th$ are parameters of the system.
To avoid using floating point arithmetic, $\Delta$ is an integer value, and we introduce another parameter to regulate the number of frames between consequent threshold updates.

\begin{figure}[!t]
\centering
\input{image/discrepance_simple_threshold.tikz}
\caption{This plot shows the ratio of frames classified for low precision computation in our technique when using a fixed threshold. To obtain this threshold, the train set is evaluated at high precision. Then, all the frames are sorted according to the number of tokens expanded during Beam Search, and the number of tokens for the frame at the required percentile is set as threshold for test\_clean and test\_other evaluation. As we can see, the desired ratio is not achieved in the test sets.}
\label{fig:discrepance_thres_simple}
\end{figure}
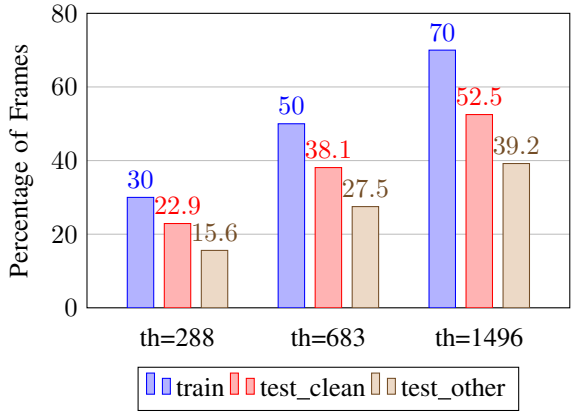

\begin{figure}[!t]
\centering
\input{image/num_tokens_cum_freq.tikz}
\caption{When low precision is used, the number of tokens expanded by Beam Search increases in general.}
\label{fig:num_tokens_cum_freq}
\end{figure}
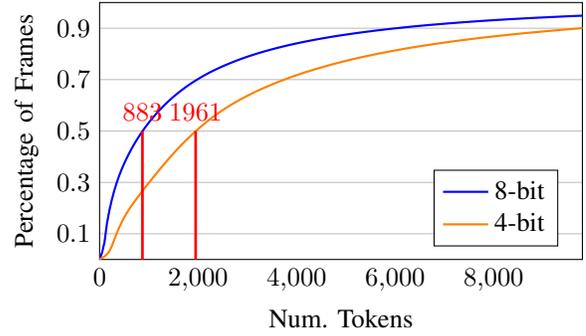

\begin{figure*}[!t]
\centering
\resizebox{2\columnwidth}{!}{
    \input{image/thresholdConvergence.tikz}
}

\caption{Threshold value computed by the proposed heuristic compared with the number of tokens expanded by Beam Search. The plot shows a span of $1.2$ million frames at $\frac{1}{1000}$ sampling rate.}
\label{fig:threshold_convergence}
\end{figure*}
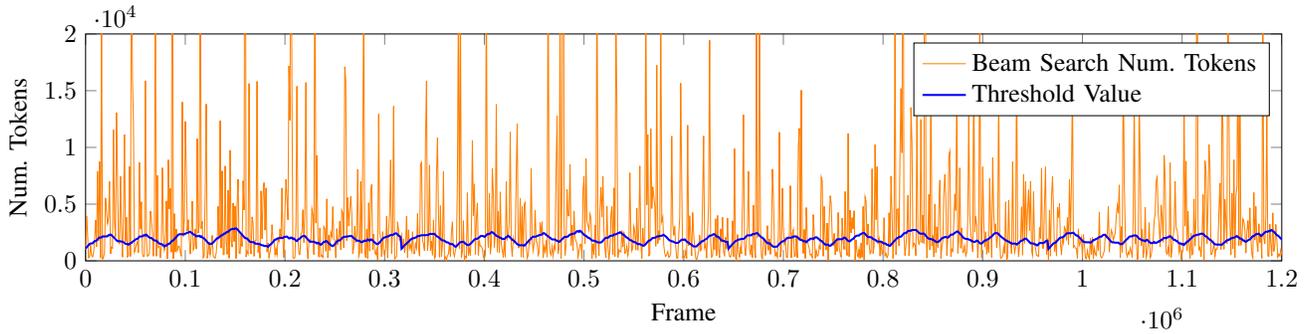

This heuristic means that the threshold is going to fluctuate. To put the amplitude of the threshold oscillations into perspective, we can compare it with the number of tokens per frame (Figure~\ref{fig:threshold_convergence}). Since the oscillations in both signals are orders of magnitude apart, we can conclude that the proposed heuristic leads to a well stable threshold.

\subsection{DNN Accelerator}

In order to implement the described technique, we quantize the weights from the full-precision Acoustic Model into two levels (8-bit and 4-bit), which we keep stored in memory. On each frame, depending on the number of tokens expanded by Beam Search, we command the DNN accelerator to evaluate the input frame using one model or the other. 

The low-precision model is half the size of the high-precision model, which results in half the time required to read the model from main memory while using the same bandwidth between the accelerator and the DRAM memory. In order to take advantage of that in the most efficient way, we modified the accelerator so it can perform computations in base precision or half precision, with the same hardware, so we can double the number of operations per cycle when operating in 4-bit mode, with very low area and power overheads over the baseline design.

The DNN accelerator now has to support two different modes of operation: \textit{base-precision} and \textit{half-precision}, which in this case means operating at $8$~bits or $4$~bits.

One of the main parameters of the accelerator is $Tn$, which configures the number of NFUs and the NFU vector size. In the baseline design, each NFU carries out the computation of a different neuron, whereas the NFU vector size enables to compute in parallel several inputs for that neuron.

Without loss of generality, we assume a configuration of the DNN accelerator with $Tn = 16$, using 8-bit weights for base-precision mode and 4-bit weights for half-precision mode. Hence, during full-precision mode, the DNN accelerator computes $16$ neurons in parallel, and for each of those, $16$ parallel inputs. In order to do so, it receives $16 \times 16$ weights and $16$ inputs each cycle.

\begin{figure}[!t]
    \centering
    \centering
    \subfloat[][]{\includegraphics{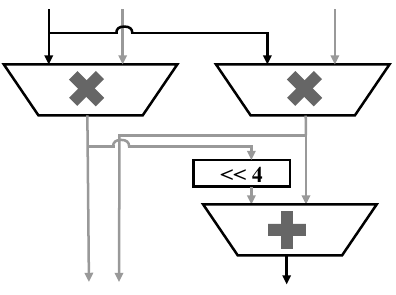}}
    \caption{Schematic of the multiplier unit included in our design. Light grey arrows represent half-precision values. This unit receives two full-precision values, which are interpreted as one full-precision and two half-precision values when operating in half-precision mode.}%
    \label{fig:mult_duplex}%
\end{figure}

For half-precision mode, however, the accelerator receives $32 \times 16$ weights and $16$ inputs per cycle (i.e., 32 neurons are computed in parallel), that is, the size of the input buffer is the same as in the baseline design. We decided not to quantize the inputs to half-precision because we found it has an important impact on WER for a small benefit on performance. During half-precision mode, we partition each compute unit so it computes $16$ inputs for $2\times16$ neurons. In this case, each compute unit would receive $2\times16$ half-precision weights and $16$ base-precision inputs. In this solution, we still have to modify the multiplication units to support both base-precision (one $8$-bit$\times8$-bit multiplication) or half-precision (two $8\times4$ multiplications). However, since in half-precision mode we are computing two neurons at the same time at each compute unit, every two multiplications share the same $8$-bit input operand. We design our multiplication units to either multiply two base-precision operands, or multiply two half-precision operands by the same base-precision operand. Figure \ref{fig:mult_duplex} shows a diagram of our multiplication unit.

\begin{figure}[!t]
    \centering
    \centering
    \subfloat[][]{\includegraphics{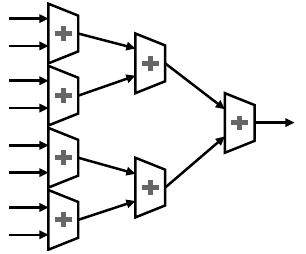}\label{fig:add_tree_simple_op}}
    \subfloat[][]{\includegraphics{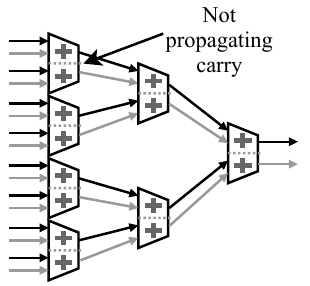}\label{fig:add_tree_duplex_op}}
    \caption{Schematic of a basic add-tree (\ref{fig:add_tree_simple_op}) and our duplex add-tree when operating in half-precision mode (\ref{fig:add_tree_duplex_op}). In the latter, each arrow represents a half-precision value.}%
    \label{fig:add_tree_duplex}%
\end{figure}

Since in half-precision mode each compute unit accumulates two different neurons, the add-tree must be able to perform one base-precision accumulation of $16$ values, or two half-precision accumulations of $16$ values each. For that purpose, we modify the adders in the tree so the transmission of the carry from one half to the other is conditioned on the mode of operation. By doing this simple modification, when the add-tree operates in half-precision mode, each adder operates as two independent adders, and thus the complete tree is unfolded in two separated trees, as shown in Figure~\ref{fig:add_tree_duplex}.

Additionally, since we are merging different levels of precision, the bit-width of the adder units has to be carefully set in order to avoid arithmetic overflow.

\subsection{DNN Activation unit and output buffer}
The activation unit performs the neuron activation function after all the neuron inputs have been accumulated. Since this unit is only used at the end of the neuron evaluation, and the accelerator is alternating the computation of several neurons (to leverage temporal locality of inputs), there is a lot of time from one activation to the next, and thus the activation unit only requires support to serialize the output from the compute unit when operating in half-precision mode. A similar argument applies for the output buffer.

\subsection{Beam Search Accelerator}
In order to compute the threshold updates following the proposed heuristic, we introduce some modifications in the Beam Search accelerator. First, it has to count the generated tokens on each search step, and keep track of the threshold, $Th$, and the variables $h$ and $h_l$ used by the heuristic. To this end, we modify the \textit{Token Issuer} to include a few adders, a very small buffer for the $h_l$ window, and a register for each variable, resulting in a negligibly area overhead.
With these modifications, the \textit{Token Issuer} keeps track of the threshold and the number of tokens expanded during each Beam Search step. After each step, the number of expanded tokens is compared with the threshold, and the required precision is exposed to the DNN accelerator. When a new frame is captured and transformed into a \textit{Feature Vector}, the DNN accelerator computes the inference for that frame in the required precision.

%% file: image/comparison_heuristics_test_clean.tikz
\begin{tikzpicture}
\begin{axis} [
    height = 5cm,
    width = 4.5cm,
    xmin = 40,
    xmax = 100,
    ymin = 6,
    ymax = 9,
    xlabel = Percentage of Frames,
    ylabel = WER,
    %width=\textwidth, 
    %height=0.6\textheight,
    legend style={
        at={(0.05,0.75)},
        anchor=west, 
        cells={anchor=west},
        fill=white, 
        fill opacity=0.6,
        draw opacity=1,
        text opacity=1
    }
]

\addplot [mark=none, blue, thick] plot coordinates {(0, 5.58) (36.1, 6.27) (50, 6.56) (67.8, 7.03) (91, 8.04) (100, 8.58)};

\addplot [mark=none, red, thick] plot coordinates {(0, 5.58) (23.7, 5.99) (40.8, 6.17) (58.1, 6.51) (65.6, 6.7) (74.15, 6.91) (100, 8.58)};

\legend{random, low\_tok}

\end{axis}
\end{tikzpicture}

%% file: image/comparison_heuristics_test_other.tikz
\begin{tikzpicture}
\begin{axis} [
    height = 5cm,
    width = 4.5cm,
    xmin = 40,
    xmax = 100,
    ymin = 16,
    ymax = 24,
    xlabel = Percentage of Frames,
    %ylabel = WER,
    %width=\textwidth, 
    %height=0.6\textheight,
    legend style={
        at={(0.05,0.75)},
        anchor=west, 
        cells={anchor=west},
        fill=white, 
        fill opacity=0.6,
        draw opacity=1,
        text opacity=1
    }
]

\addplot [mark=none, blue, thick] plot coordinates {(0, 15.7) (35.8, 16.61) (50, 17.5) (67.45, 18.9) (100, 23.15)};

\addplot [mark=none, red, thick] plot coordinates {(0, 15.7) (32.88, 16.49) (51.12, 17.05) (62.66, 17.86) (70.1, 18.37) (76.9, 19.2) (100, 23.15)};

\legend{random, low\_tok}

\end{axis}
\end{tikzpicture}

%% file: image/discrepance_simple_threshold.tikz
% Preamble: \pgfplotsset{width=7cm,compat=1.5.1}
\begin{tikzpicture}
    \begin{axis}[
    height=5.5cm,
    width=8cm,
    ymax=80,
    major x tick style = transparent,
    ybar=4pt,
    enlarge x limits=0.3,
    bar width=10pt,
    ymajorgrids = true,
    ylabel={Percentage of Frames},
    symbolic x coords={th=288, th=683, th=1496},
    xtick = data,
    scaled y ticks = false,
    ymin=0,
    ytick style={draw=none},
    legend cell align=left,
    legend style={
        at={(0.5,-0.2)},
        anchor=north,
        legend columns=-1
    },
    nodes near coords,
    nodes near coords align={vertical},
    ]
        
        \addplot coordinates {(th=288,30) (th=683,50) (th=1496,70)};
        \addplot coordinates {(th=288,22.9) (th=683,38.1) (th=1496,52.5)};
        \addplot coordinates {(th=288,15.6) (th=683,27.5) (th=1496,39.2)};
        
        \legend{train, test\_clean,test\_other}
        
    \end{axis}
\end{tikzpicture}

%% file: image/num_tokens_cum_freq.tikz
\begin{tikzpicture}
\begin{axis} [
    height=5cm,
    width=8cm,
    xmin = 0,
    xmax = 9801,
    ymin = 0,
    ymax = 1,
    xlabel = Num. Tokens,
    ylabel = Percentage of Frames,
    major x tick style = transparent,
    ytick style={draw=none},
    ymajorgrids = true,
    ytick={0.1, 0.3, 0.5, 0.7, 0.9},
    %width=\textwidth, 
    %height=0.6\textheight,
    legend style={at={(0.95,0.2)},anchor=east, cells={anchor=west}}
]

\addplot [mark=none, blue, thick] table [x=binLimit, y=cumFreqHP, col sep=comma] {data/num_token_cum_freq.txt};
\addplot [mark=none, orange, thick] table [x=binLimit, y=cumFreqLP, col sep=comma] {data/num_token_cum_freq.txt};

\draw [line width=1pt, red] (195,0) -- (195,50) node [above] {$1961$};
\draw [line width=1pt, red] (87,0) -- (87,50) node [above] {$883$};

\legend{8-bit, 4-bit}

\end{axis}
\end{tikzpicture}

%% file: image/thresholdConvergence.tikz
\begin{tikzpicture}
\begin{axis} [
    xmin = 0,
    xmax = 1200000,
    ymin = 0,
    ymax = 20000,
    xlabel = Frame,
    ylabel = Num. Tokens,
    width=\textwidth, 
    height=.2\textheight,
    legend style={at={(0.99,0.8)},anchor=east, cells={anchor=west}}
]
\addplot [mark=none, orange] table [x=frame, y=numTokens, col sep=comma] {data/thres_value_trace.txt};
\addlegendentry{Beam Search Num. Tokens}

\addplot [mark=none, blue, thick] table [x=frame, y=threshold, col sep=comma] {data/thres_value_trace.txt};
\addlegendentry{Threshold Value}

\end{axis}
\end{tikzpicture}

%% file: text/7_methodology.tex
\section{methodology} \label{sec:methodology}

To measure performance and energy consumption of the accelerators, we relied on cycle-accurate simulators to count cycles and usage of logic units. We model the mobile SoC platform described in Section~\ref{sec:asr_hardware}, illustrated in Figure~\ref{fig:asr_pipeline}. The hardware paremeters for the experiments are shown in Table~\ref{tbl:table_soc}. To obtain the power values for all the logic units, we modeled them in Verilog, employing IPs from the  \textit{Synopsys DesignWare Building Block IP} when available, and writing the Verilog ourselves when required. To report accurate power, we employed the \textit{Synopsis Power Compiler} tool, which requires a \textit{switching activity} file. These activity files were obtained by simulating the Verilog designs with specific test-benches.

The SRAM memories were modeled through the modified version of \textit{CACTI}~\cite{muralimanohar2009cacti} included in \textit{McPAT}~\cite{li2009mcpat}, whereas the DRAM memory was modeled with \textit{Micron DRAM Power model}~\cite{micron2019power}. The later requires the memory usage conditions, which we set by estimating the average read and write bandwidth required by the accelerators when operating at different frequencies.

The clock frequency for each logic unit was derived from the critical path as estimated by the \textit{Synopsis Synthesis tool} from our Verilog code and the limit established by the bandwidth requirement between the accelerators and the main memory. The clock frequency for the Beam Search accelerator is constrained by the logic units, and was set to $600$~Mhz whereas the frequency for the DNN accelerator is limited by the main memory bandwidth and was set to $55$~Mhz. Of course, a DRAM with higher bandwidth could be used to improve the performance, but that would increase the overall system's power and energy consumption, rendering it less attractive for low-power solutions.

To measure the CPU performance, we used a \textit{Jetson Tx1}~Board from \textit{Nvidia}, which contains, among other components, the ARM CPU that we are assuming for our baseline. Those tasks previously identified to run on CPU were executed on this board, while reading the internal performance counters for time and CPU energy. To ensure that the measurements were accurate, we launched the utterance evaluations one by one, making sure that the GPU was not being used, and that this process was the only one running (apart from OS tasks).

Regarding the speech corpus, we employed Librispeech~\cite{panayotov2015librispeech}, which is composed of 1000 hours of speech, divided in 5 sets: train, test\_clean, test\_other, dev\_clean and dev\_other. The train set, augmented with different techniques included in Kaldi, was used to train the models (see Table~\ref{tbl:table_asr}), whereas test\_clean and test\_other, containing more than 5 hours of speech each, were used for the evaluations. The difference between both test sets is that test\_other contains more challenging utterances than test\_clean.

%% file: text/6_experimental_results.tex
\section{Experimental Results} \label{sec:experimental_results}

In this section we evaluate the speedups and energy savings achieved by our dynamic DNN precision scheme based on Beam Search confidence. The baseline system is the mobile SoC platform described in Section~\ref{sec:asr_hardware}. It includes a multicore
ARM CPU, a DNN accelerator and a Beam Search accelerator. We have implemented our scheme on top of this SoC
as described in Section~\ref{sec:platform}.

In order to implement our technique, the baseline accelerators require some modifications. More specifically, the DNN accelerator must support two operation modes: base-precision and half-precision, whereas the Beam Search accelerator has to compute the heuristic and maintain the threshold. These modifications, made as described in Section \ref{sec:platform}, result on an area overhead of $3.1$\% over the baseline accelerators, resulting on a negligible overhead over the complete platform, since the accelerators are very small when compared to the 8GB~LPDDR4 or the 4-core ARM CPU. The DNN accelerator occupies an area of $0.42 mm^2$, split between \textit{buffers} ($25.5\%$), \textit{MULT arrays} ($62.6\%$) and \textit{AddTrees} ($11.9\%$), whereas the Beam Search accelerator occupies $3.34 mm^2$.

Since we keep on memory an additional DNN model (AM quantized to 4 bits), our solution incurs on memory footprint overheads. However, since the additional model is fairly small, the overall overhead is just $3.8$\% increase in memory footprint.

The average power of the complete system, including CPU, accelerators and external DRAM, is $1.17W$ ($3.3$\% increase over the baseline), and the frames are evaluated 20x faster than real time, consuming $0.7$ mJ/frame.

\subsection{Performance Gains of Dynamic Precision AM}

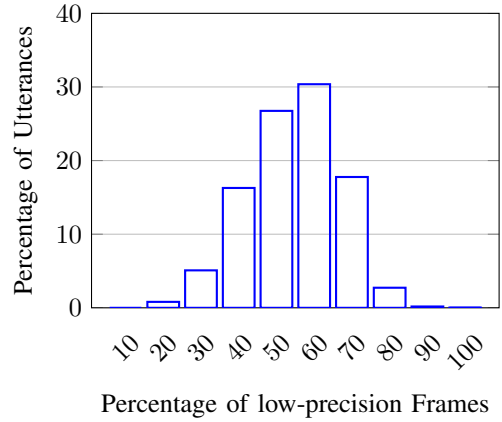
\begin{figure}[!t]
\centering
\input{image/ratio_lp_freq.tikz}
\caption{Frequency of utterances (vertical axis) grouped by the percentage of their frames computed at low precision (horizontal axis).}
\label{fig:ratio_frames_lp}
\end{figure}

Since our heuristic modifies the threshold slowly, during a single utterance evaluation it does not change by a large extent, and thus, although for a long run the number of frames evaluated in low precision converges at $50$\%, that is not the case for individual utterances. Figure~\ref{fig:ratio_frames_lp} shows the distribution of utterances according to the percentage of frames evaluated at low precision. Most of them fall between $40-60$\%, but a few of them evaluated at low precision more than $80$\% of their frames.

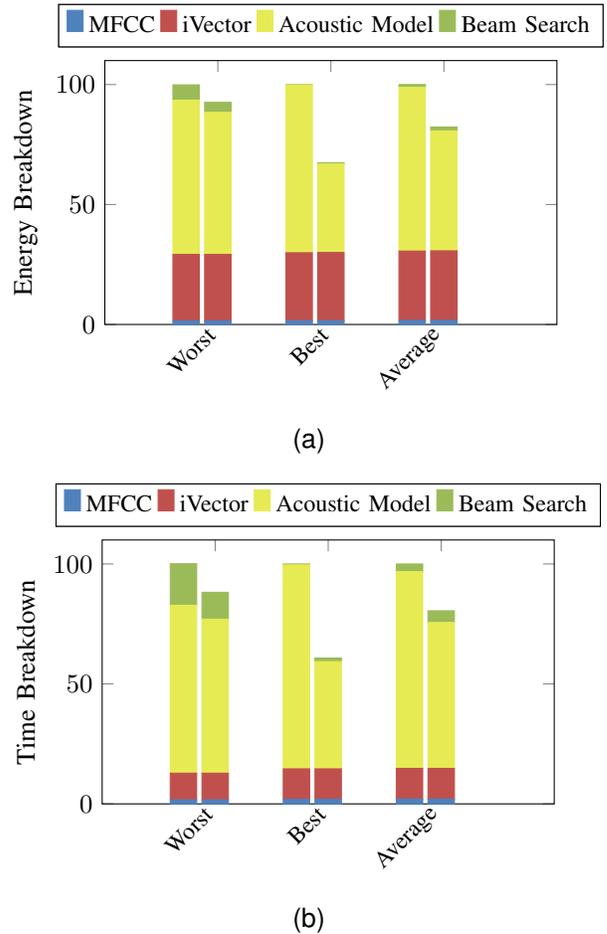
\begin{figure}[!t]
    \centering
    \centering
    \subfloat[][]{\input{image/result_energy_breakdown.tikz}\label{fig:result_energy_breakdown_ASR}}
    
    \subfloat[][]{\input{image/result_time_breakdown.tikz}\label{fig:result_time_breakdown_DNN}}
    \caption{Energy consumption (\textit{a}) and execution time (\textit{b}) normalized to the baseline. The bars in each plot represent: \textit{Worst Utterance}, \textit{Best Utterance} and \textit{Test Set Average} for the baseline and proposed scheme respectively.}%
    \label{fig:result_breakdown}%
\end{figure}

Figure~\ref{fig:result_breakdown} shows the savings obtained from the proposed technique compared to the evaluation on the baseline platform. Both for energy and time, three cases are plotted: the \textit{Worst Utterance}, the \textit{Best Utterance} and the \textit{Test Set Average}. The \textit{Worst} and \textit{Best} utterances are chosen regarding the percentage of frames evaluated at half-precision. We can see how even for the worst case, significant savings are achieved.

By applying our technique, we can save up to $47.2$\% of energy and reduce the execution time up to $47.4$\% for the Acoustic Model evaluation on utterances where the Beam Search confidence is high for most of its frames. On average, our scheme reduces energy consumption by $25.6$\% and execution time by $25.8$\% when evaluating the complete test set.

Since the low-precision ($4$-bit weights) Acoustic Model network is half the size than the full-precision ($8$-bit weights) network, whenever a frame is evaluated at low-precision, we save half the reads from main memory. Operating at half-precision results in significant speedups for two reasons. First, the Neural Function Units (NFUs) are modified so they can operate at double throughput in half-precision mode with negligible hardware overheads. Second, the DNN accelerator is memory bound since data reuse is largely limited in TDNN networks and, hence, reducing one half of the reads from main memory results in large performance improvements. Therefore, the execution time for Acoustic Model evaluation is reduced by approximately one half during low-precision frames.

On the other hand, the reduction in energy consumption is also mostly explained by the reduction in reads from off-chip memory. As detailed in Section~\ref{sec:analysis}, off-chip reads of the acoustic model weights are the main bottleneck of the system, contributing to $85$\% of the energy consumed during Acoustic Model evaluation. Another source of energy savings comes from the reduction in static energy consumed by the rest of the components during the time that the Acoustic Model is being evaluated. Since around $50$\% of the total number of frames are evaluated at low precision, the observed savings of around $25$\% in time and energy during Acoustic Model evaluation are consistent.

When we take into account the complete ASR system, the savings obtained in the acoustic model evaluation translate to an average reduction on energy consumption of $16.9$\% and a reduction of execution time of $19.5$\% (Figure~\ref{fig:result_breakdown}). As discussed in Section~\ref{sec:analysis}, when the low precision acoustic model is employed, the average confidence of the Beam Search is decreased, which translates to a decrease in the performance of the Beam Search when our technique is used. Consequently, the energy and time consumed by the Beam Search is generally increased with respect to the baseline. However, even for the worst cases observed in the test sets, the benefits outperform the overheads, resulting in a net improvement in performance for all the utterances.

\subsection{Effect on Accuracy}

As discussed in Section~\ref{sec:analysis}, reducing the accuracy for the Acoustic Model evaluation results in a minor degradation in the recognition accuracy when the frames for low precision evaluation are carefully chosen. Our experiments show that by using the proposed heuristic based on the number of tokens, we can compute at low precision $50$\% of the test sets frames incurring in less than $1$\% absolute WER loss for test\_clean, and $2.69$\% for test\_other. 

\begin{figure}[!t]
    \centering
    \centering
    \subfloat[][]{\input{image/sens_analysis_test_clean.tikz}}
    %\qquad
    \subfloat[][]{\input{image/sens_analysis_test_other.tikz}}
    \caption{Sensitivity analysis of the percentage of frames set as target for low precision evaluation. The curves represent the WER obtained when some percentage of frames is evaluated in low precision for test\_clean (a), and test\_other (b).}
\label{fig:sensAnalysis}%
\end{figure}
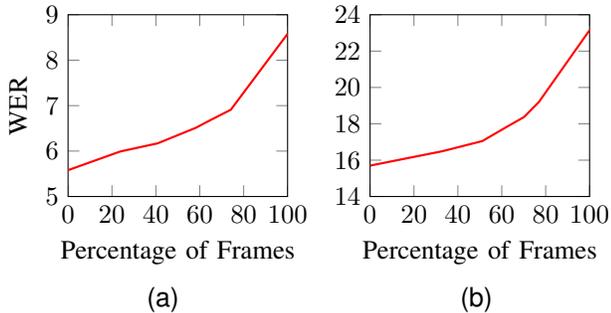

In order to select a target for the percentage of frames computed at low precision, we performed a sensitivity analysis, modifying the target percentage from $0$\% (every frame in high precision) to $100$\% (every frame in low precision).
Figure~\mbox{\ref{fig:sensAnalysis}} shows the relation between WER and savings. Note that the time and energy savings are proportional to the percentage of frames evaluated at low precision. If some percentage of frames, $x$\%, is evaluated in low precision, we save around $x/2$\% of time and energy during the DNN evaluation. The figure shows a curve with an elbow around $x = 50$\% for both cases: test\_clean, and test\_other. Choosing this target grants a significant gain with negligible WER loss.

%% file: image/ratio_lp_freq.tikz
% Preamble: \pgfplotsset{width=7cm,compat=1.5.1}
\begin{tikzpicture}
    \begin{axis}[
        height = 5.5cm,
        width = 7cm,
        major x tick style = transparent,
        ybar=0.05pt,
        bar width=12pt,
        ymajorgrids = true,
        xtick = data,
        ylabel={Percentage of Utterances},
        xlabel={Percentage of low-precision Frames},
        scaled y ticks = false,
        ymin=0,
        ymax=40,
        x tick label style={rotate=45}
        %xmin=10,
        %xmax=100
    ]
    
        \addplot [mark=none, blue, thick] table [x=ratioSmall, y=percentUtts, col sep=comma] {data/ratio_lp_freq.txt};
    
    \end{axis}
\end{tikzpicture}

%% file: image/result_energy_breakdown.tikz
\begin{tikzpicture}

\definecolor{bblue}{HTML}{4F81BD}
\definecolor{rred}{HTML}{C0504D}
\definecolor{ggreen}{HTML}{9BBB59}
\definecolor{ppurple}{HTML}{9F4C7C}
\definecolor{yyellow}{HTML}{E6EB53}
\definecolor{lightgray}{gray}{0.9}

\definecolor{rosso}{RGB}{220,57,18}
\definecolor{giallo}{RGB}{255,153,0}
\definecolor{blu}{RGB}{102,140,217}
\definecolor{verde}{RGB}{16,150,24}
\definecolor{viola}{RGB}{153,0,153}

\pgfplotsset{
    every axis/.style={ % add these settings to all the axis environments in the tikzpicture
    height=\axisdefaultheight*0.7,
    width=\axisdefaultwidth*0.9,
    ybar stacked,
    enlarge x limits=0.5,
    ymin=0,ymax=110,
    x tick label style={
        rotate=45,
        anchor=east,
        align=right,
        font=\small,
        text width=1.2cm
        },
    symbolic x coords={Worst,Best,Average},
    xtick=data,
    bar width=10pt,
    ylabel={Energy Breakdown},
    legend cell align=left,
    legend style={
        at={(0.5,1.05)},
        anchor=south,
        legend columns=-1,
        font=\small
        }
  }}

% BASELINE
\begin{axis}[bar shift=-12pt]
\addplot[style={bblue,fill=bblue,mark=none}] % MFCC
  coordinates {(Worst,1.5) (Best,1.6) (Average,1.7)};
\addplot[style={rred,fill=rred,mark=none}]   % IVECTOR
  coordinates {(Worst,27.7) (Best,28.3) (Average,28.9)};
\addplot[style={yyellow,fill=yyellow,mark=none}] % AM
  coordinates {(Worst,64.3) (Best,69.9) (Average,68.3)};
\addplot[style={ggreen,fill=ggreen,mark=none}] % BS
  coordinates {(Worst,6.4) (Best,0.1) (Average,1.1)};
\legend{MFCC, iVector, Acoustic Model, Beam Search}
\end{axis}

% PROPOSAL
\begin{axis}[bar shift=+0pt,hide axis]
\addplot[style={bblue,fill=bblue,mark=none}] % MFCC
  coordinates {(Worst,1.5) (Best,1.6) (Average,1.7)};
\addplot[style={rred,fill=rred,mark=none}]   % IVECTOR
  coordinates {(Worst,27.7) (Best,28.4) (Average,29)};
\addplot[style={yyellow,fill=yyellow,mark=none}] % AM
  coordinates {(Worst,59.2) (Best,36.9) (Average,50)};
\addplot[style={ggreen,fill=ggreen,mark=none}] % BS
  coordinates {(Worst,4.2) (Best,0.5) (Average,1.6)};
\legend{MFCC, iVector, Acoustic Model, Beam Search}
\end{axis}

\end{tikzpicture}

%% file: image/result_time_breakdown.tikz
\begin{tikzpicture}

\definecolor{bblue}{HTML}{4F81BD}
\definecolor{rred}{HTML}{C0504D}
\definecolor{ggreen}{HTML}{9BBB59}
\definecolor{ppurple}{HTML}{9F4C7C}
\definecolor{yyellow}{HTML}{E6EB53}
\definecolor{lightgray}{gray}{0.9}

\definecolor{rosso}{RGB}{220,57,18}
\definecolor{giallo}{RGB}{255,153,0}
\definecolor{blu}{RGB}{102,140,217}
\definecolor{verde}{RGB}{16,150,24}
\definecolor{viola}{RGB}{153,0,153}

\pgfplotsset{
    every axis/.style={ % add these settings to all the axis environments in the tikzpicture
    height=\axisdefaultheight*0.7,
    width=\axisdefaultwidth*0.9,
    ybar stacked,
    enlarge x limits=0.5,
    ymin=0,ymax=110,
    x tick label style={
        rotate=45,
        anchor=east,
        align=right,
        font=\small,
        text width=1.2cm
        },
    symbolic x coords={Worst,Best,Average},
    xtick=data,
    bar width=10pt,
    ylabel={Time Breakdown},
    legend cell align=left,
    legend style={
        at={(0.5,1.05)},
        anchor=south,
        legend columns=-1,
        font=\small
        }
  }}

% BASELINE
\begin{axis}[bar shift=-12pt]
\addplot[style={bblue,fill=bblue,mark=none}] % MFCC
  coordinates {(Worst,1.7) (Best,1.9) (Average,2)};
\addplot[style={rred,fill=rred,mark=none}]   % IVECTOR
  coordinates {(Worst,11.1) (Best,12.7) (Average,12.8)};
\addplot[style={yyellow,fill=yyellow,mark=none}] % AM
  coordinates {(Worst,70) (Best,85) (Average,82)};
\addplot[style={ggreen,fill=ggreen,mark=none}] % BS
  coordinates {(Worst,17.3) (Best,0.4) (Average,3.2)};
\legend{MFCC, iVector, Acoustic Model, Beam Search}
\end{axis}

% PROPOSAL
\begin{axis}[bar shift=+0pt,hide axis]
\addplot[style={bblue,fill=bblue,mark=none}] % MFCC
  coordinates {(Worst,1.7) (Best,1.9) (Average,2)};
\addplot[style={rred,fill=rred,mark=none}]   % IVECTOR
  coordinates {(Worst,11.1) (Best,12.7) (Average,12.8)};
\addplot[style={yyellow,fill=yyellow,mark=none}] % AM
  coordinates {(Worst,64.2) (Best,44.7) (Average,60.8)};
\addplot[style={ggreen,fill=ggreen,mark=none}] % BS
  coordinates {(Worst,11.2) (Best,1.5) (Average,4.9)};
\legend{MFCC, iVector, Acoustic Model, Beam Search}
\end{axis}

\end{tikzpicture}

%% file: image/sens_analysis_test_clean.tikz
\begin{tikzpicture}
\begin{axis} [
    height = 4cm,
    width = 4.5cm,
    xmin = 0,
    xmax = 100,
    ymin = 5,
    ymax = 9,
    xlabel = Percentage of Frames,
    ylabel = WER,
    %width=\textwidth, 
    %height=0.6\textheight,
    legend style={
        at={(0.05,0.75)},
        anchor=west, 
        cells={anchor=west},
        fill=white, 
        fill opacity=0.6,
        draw opacity=1,
        text opacity=1
    }
]

\addplot [mark=none, red, thick] plot coordinates {(0, 5.58) (23.7, 5.99) (40.8, 6.17) (58.1, 6.51) (65.6, 6.7) (74.15, 6.91) (100, 8.58)};

\end{axis}
\end{tikzpicture}

%% file: image/sens_analysis_test_other.tikz
\begin{tikzpicture}
\begin{axis} [
    height = 4cm,
    width = 4.5cm,
    xmin = 0,
    xmax = 100,
    ymin = 14,
    ymax = 24,
    xlabel = Percentage of Frames,
    %ylabel = WER,
    %width=\textwidth, 
    %height=0.6\textheight,
    legend style={
        at={(0.05,0.75)},
        anchor=west, 
        cells={anchor=west},
        fill=white, 
        fill opacity=0.6,
        draw opacity=1,
        text opacity=1
    }
]

\addplot [mark=none, red, thick] plot coordinates {(0, 15.7) (32.88, 16.49) (51.12, 17.05) (62.66, 17.86) (70.1, 18.37) (76.9, 19.2) (100, 23.15)};

\end{axis}
\end{tikzpicture}

%% file: text/8_related_work.tex
\section{Related Work} \label{sec:related_work}

The field of Machine Learning has experienced an enormous growth in the last few decades, and that has been reflected in the computer architecture research, with a high number of proposals to increase the performance of Machine Learning algorithms. Since heterogeneous systems have also been gaining popularity, many of the new proposals regarding machine learning are related to accelerators for specific algorithms, being DNN acceleration a popular alternative. However, many of the proposed accelerators focus on exploiting characteristics of specific types of networks, such as CNNs~\cite{du2015shidiannao, chen2016eyeriss, chen2019eyeriss} or RNNs~\cite{silfa2018pur, gupta2019masr}, which are not a good fit for our baseline ASR system, which relies on a TDNN acoustic model~\cite{povey2018time, peddinti2015time}, mostly composed of fully-connected layers. Other recent works focus on high performance computing. Chen et al.~\cite{luo2016dadiannao} propose \textit{DaDianNao}, a modular accelerator for DNN and CNN composed of $67.72mm^2$ nodes, consuming $15.97W$ each. Song et al.~\cite{han2017ese} designed \textit{ESE}, a DNN accelerator optimized for LSTM layers, consuming $41W$ on an FPGA. This work is different because we focus on low-power on-edge computation, with tight constraints in area and power dissipation.

Regarding specific proposals for speech recognition, prior work was focused on older ASR systems, and assumed smaller vocabularies and/or acoustic models than state-of-the-art solutions as the one considered in this paper. Price et al.~\cite{price2016energy} designed a chip encompassing from \textit{Voice Activity Detection} (\textit{VAC}) and audio capture, to Beam Search decoding. The area of the chip is $13.18 mm^2$, and consumes $11.27 mW$ (not including power from off-chip components, such as main memory, which is the main bottleneck according to our models) while running a $145k$ word vocabulary benchmark. On the other hand, our work focus on bigger models, including a $200k$ vocabulary decoding graph, and a $16.17MB$ acoustic model (opposed to their $3.71MB$ model) to achieve state-of-the-art accuracy. 

Other proposals try to optimize a specific part using only information local to that component, for example, Yazdani et al.~\cite{yazdani2017low, yazdani2016ultra} propose a Beam Search accelerator, consuming $462 mW$ plus the power dissipated by the GPU (between $2W$ and $6W$), which is used for the Acoustic Model evaluation. Additionally, they optimize the Beam Search accelerator by performing on-the-fly composition of the WFST-based decoding graph~\cite{yazdani2017unfold}. Our work is different since our aim is not to improve the Beam Search energy-efficiency, but to leverage information known in the Beam Search accelerator to improve energy and performance of the DNN accelerator. 

Another common optimization consists in exploiting computation reuse in the DNN accelerator~\cite{riera2018computation, jiao2018energy, ning2019deep, hegde2018ucnn}, or compressing the DNN model, e.g. via weight pruning~\cite{han2017ese, gupta2019masr}. In this work, we follow a different approach. Instead of optimizing a component by exploiting local properties, we look at the system from a high-level perspective, exploiting the inter-dependencies among ASR stages. Note, however, that our proposal can still be applied on top of any of these optimizations, providing additional gains.

Earlier proposals for hardware accelerated ASR~\cite{miura2008low, lin2006moving, chun2011isis} focused on GMM based recognizers, with \textit{CMU's Sphinx} as an usual software baseline, and vocabularies with less than 100k words (e.g. 5k/20k-word Wall Street Journal, 64K-word Broadcast News,...). More recently, Tabani et. al.~\cite{tabani2017ultra} proposed an accelerator for the \textit{PocketSphinx} system, configured to decode a 130k-word librispeech-based benchmark. \textit{PocketSphinx} is based on \textit{CMU Sphinx}, aimed at portability. By using that accelerator (a $0.94mm^2$, $110mW$ chip), the decoding time and energy is reduced by $5.89x$ and $241x$, respectively, over a mobile GPU implementation. However, this type of systems have become less popular nowadays due to their lower accuracy. For instance, Tabani et. al.~\cite{tabani2017ultra} report a WER of $24.14$, which is much higher than current state-of-the-art systems.

%% file: text/9_conclusions.tex
\section{Conclusions} \label{sec:conclusions}

In this work, we show how the number of tokens expanded during the Beam Search can be used to improve the performance of the Acoustic Model evaluation, the main bottleneck of ASR systems.

Following the observation that a low number of expanded tokens is related with high confidence in the partial decoding, we set a threshold on the number of tokens and evaluate in low precision the Acoustic Model for those frames falling below it. This threshold is computed in run-time by using a heuristic that guarantees that $50$\% of the frames will be computed at low precision.

To support this computation scheme, we modified a baseline low-power DNN accelerator, changing the array multiplication and add-tree units by specifically designed duplex units, so the accelerator can operate either in base or half precision, doubling the throughput for the low precision frames, with minimal impact on area and power.
By using our proposal, the performance of the DNN-based Acoustic Model evaluation is improved by $25.9$\%, whereas the energy consumption is reduced by $25.6$\% on average for the entire Librispeech test set.

%% file: text/_1_acknowledgments.tex
\section*{Acknowledgements}
This work has been supported by the CoCoUnit ERC Advanced Grant of the EU’s Horizon 2020 program (grant No 833057), the Spanish State Research Agency under grant TIN2016-75344-R (AEI/FEDER, EU), the ICREA Academia program and the Spanish MICINN Ministry under grant BES-2017-080605.